\documentclass[twocolumn,showpacs,preprintnumbers,amsmath,amssymb,superscriptaddress]{revtex4}

\usepackage{graphicx}
\begin{document}
\bibliographystyle{prsty}
\title{Electronic charges and electric potential at LaAlO$_3$/SrTiO$_3$ interfaces studied by core-level photoemission spectroscopy}

\author{M.~Takizawa}
\affiliation{Department of Physics, University of Tokyo, 
3-7-1 Hongo, Bunkyo-ku, Tokyo, 113-0033, Japan}
\author{S.~Tsuda}
\affiliation{Department of Advanced Materials Science, University of Tokyo,  
5-1-5 Kashiwanoha, Kashiwashi, Chiba, 277-8651, Japan}
\author{T.~Susaki}
\affiliation{Department of Advanced Materials Science, University of Tokyo,  
5-1-5 Kashiwanoha, Kashiwashi, Chiba, 277-8651, Japan}
\author{H.~Y.~Hwang}
\affiliation{Department of Advanced Materials Science, University of Tokyo,  
5-1-5 Kashiwanoha, Kashiwashi, Chiba, 277-8651, Japan}
\affiliation{Japan Science and Technology Agency, Kawaguchi, 332-0012, Japan}
\affiliation{Department of Applied Physics and Stanford Institute for Materials and Energy Science, Stanford University, Stanford, California 94305, USA}
\author{A.~Fujimori}
\affiliation{Department of Physics, University of Tokyo, 
3-7-1 Hongo, Bunkyo-ku, Tokyo, 113-0033, Japan}

\date{\today}

\begin{abstract}
We studied LaAlO$_{3}$/SrTiO$_{3}$ interfaces for varying LaAlO$_{3}$ thickness by core-level photoemission spectroscopy. 
In Ti $2p$ spectra for conducting ``$n$-type'' interfaces, Ti$^{3+}$ signals appeared, which were absent for insulating ``$p$-type'' interfaces. 
The Ti$^{3+}$ signals increased with LaAlO$_{3}$ thickness, but started well below the critical thickness of 4 unit cells for metallic transport. 
Core-level shifts with LaAlO$_{3}$ thickness were much smaller than predicted by the polar catastrophe model. 
We attribute these observations to surface defects/adsorbates providing charges to the interface even below the critical thickness. 
\end{abstract}


\maketitle
Recently, heterostructures consisting of two different metal oxides have attracted considerable interest because of their fundamental interest as well as their possible device applications. 
At the interfaces between two band insulators, LaAlO$_{3}$ and SrTiO$_3$, metallic conduction with high carrier mobility \cite{OhtomoLAOSTO} and even superconductivity \cite{LAOSTO-super} have been observed. 
The transport at the LaAlO$_3$/SrTiO$_3$ interfaces shows a remarkable termination-layer dependence \cite{OhtomoLAOSTO}. 
That is, the (LaO)$^{+}$/(TiO$_{2}$)$^{0}$ interfaces exhibit metallic conductivity while the (AlO$_{2}$)$^{-}$/(SrO)$^{0}$ interfaces remain insulating. 
Moreover, the metallic transport at the LaAlO$_3$/SrTiO$_3$ interfaces occurs only beyond a critical LaAlO$_3$ layer thickness of 4-6 unit cells (uc), i.e., 1.6-2.3 nm \cite{HuijbenLAOSTO, ThielLAOSTO}. 

In order to interpret such properties of the interfaces, an electronic reconstruction which avoids a ``polar catastrophe'' has been proposed \cite{polarReview, HesperK3C60, NakagawaLAOSTO, Lee-calc}. 
Because the (001) planes of LaAlO$_3$ are polar, that is, positively charged (LaO)$^{+}$ and negatively charged (AlO$_{2}$)$^{-}$ are alternatingly stacked, the electrostatic potential difference between both surfaces of the LaAlO$_3$(001) thin film should increase proportional to the thickness of the LaAlO$_3$ layer. 
To avoid such an energetically unfavorable divergence of electrostatic potential, some charge redistribution or electronic reconstruction should take place. 
For the (LaO)$^{+}$/(TiO$_{2}$)$^{0}$ interfaces (``$n$-type'' interfaces), the polar catastrophe can be avoided if half an electron is added to the interfacial region, resulting in the conversion of the valence of interface Ti atoms from the original Ti$^{4+}$ to Ti$^{4+}$-Ti$^{3+}$ mixed valence. 
For the (AlO$_{2}$)$^{-}$/(SrO)$^{0}$ interfaces (``$p$-type'' interfaces), the divergence can be avoided if half an electron is removed from the interface through hole doping into the O $2p$ band or through the formation of oxygen vacancies. 
On the other hand, some studies \cite{Kalabukhov, Siemons, Herranz} have suggested that the interfacial conductivity originated from the formation of oxygen vacancies in the entire region of the SrTiO$_{3}$. 
A cross-sectional conducting-tip atomic force microscope study of LaAlO$_3$/SrTiO$_3$ interfaces \cite{LAOSTO-AFM} has revealed that the high-mobility electron gas is confined within a few nanometers when annealed in oxygen atmosphere.  
A recent study has shown that not only the carrier density but also the carrier mobility is a strong function of the LaAlO$_3$ layer thickness \cite{Bell}. 
Indeed, the generally low carrier concentration deduced from transport measurements compared to that necessary to avoid the polar catastrophe, as well as recent experimental \cite{Dikin} and theoretical \cite{Pentcheva-calc3, Popovic-cal2} studies, have pointed to major effects of carrier localization at the interfaces even above the critical LaAlO$_{3}$-layer thickness. 
Thus the origin of the metallic interfacial conductivity still remains highly controversial and is becoming even more non-trivial. 

Photoemission spectroscopy (PES), which has been extensively used to study surface and bulk properties of solids, has been found to be a powerful technique to investigate oxide interfaces as well \cite{TakizawaLTOSTO, WadatiLVO, LAOSTO-HXPES, LAOSTO-Yoshimatsu}. 
A hard x-ray PES study of the Ti core level \cite{LAOSTO-HXPES} has shown that a few percent of Ti$^{3+}$ exist within a few nanometers of the $n$-type LaAlO$_3$/SrTiO$_3$ interface, indicating the electronic reconstruction. 
However, according to an x-ray photoemission (XPS) study of both $n$-type and $p$-type LaAlO$_3$/SrTiO$_3$ interfaces \cite{Segal}, the core-level shifts were much smaller than and opposite to those predicted by the polar catastrophe model. 
Soft x-ray PES studies of the valence band and core levels have also been performed on the same systems by Yoshimatsu {\it et al} \cite{LAOSTO-Yoshimatsu}. 
They deduced the band bending on the SrTiO$_3$ side of the interface from the Ti $2p$ core-level shift with varying LaAlO$_3$ thickness, and attributed the metallic conductivity to electrons trapped by the triangular potential at the interface, although they could not detect electrons which should be doped into the Ti $3d$ band at the LaAlO$_3$/SrTiO$_3$ interface. 
In the present work, in order to resolve the origins of the apparently conflicting results and clarify the electronic structure of the LaAlO$_3$/SrTiO$_3$ interface, we have performed core-level XPS measurements with varying LaAlO$_3$ thickness both for $n$-type and $p$-type interfaces. 
The results showed a systematic increase of the Ti$^{3+}$ component in the Ti $2p$ core-level spectra with LaAlO$_3$-layer thickness only for $n$-type interfaces, consistent with the electronic reconstruction, whereas there was no abrupt increase at the critical LaAlO$_3$-layer thickness of $\sim$ 4 uc, and no clear core-level shifts for $p$-type interfaces. 
In order to interpret the experimental results, we suggest that the influence of charged surface defects has to be incorporated. 

\begin{figure}
\begin{center}
\includegraphics[width=\linewidth]{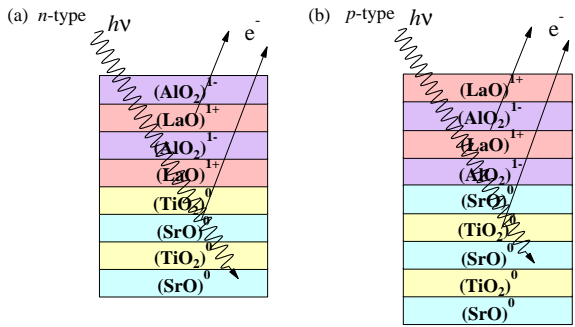}
\caption{(Color online) Schematic view of LaAlO$_{3}$ thin films grown on SrTiO$_{3}$ (001) substrates and their photoemission measurement geometry. Panels (a) and (b) represent the two types of samples assuming bulk-like charge assignments. (a) LaAlO$_3$ is deposited on a TiO$_2$-terminated SrTiO$_3$ substrate forming the ``$n$-type'' interface. (b) LaAlO$_3$ is deposited on a SrO-terminated SrTiO$_3$ substrate forming the ``$p$-type'' interface. }
\label{sample}
\end{center}
\end{figure}
LaAlO$_3$ thin films with varying thickness from 1 to 6 uc were grown on the atomically flat, TiO$_{2}$-terminated (001) surfaces of SrTiO$_3$ substrates using pulsed laser deposition (PLD). 
As schematically shown in Fig.~\ref{sample}(a), the $n$-type interface is formed when the LaAlO$_3$ thin film is deposited onto this TiO$_{2}$-terminated surface of the SrTiO$_3$ substrate. 
The $p$-type interface is formed when the LaAlO$_3$ thin film is deposited onto the SrO-terminated surface of the SrTiO$_3$ substrate [Fig.~\ref{sample}(b)]. 
Here, the TiO$_2$-terminated surface of the SrTiO$_3$ (001) substrates could be converted to a SrO-terminated surface by depositing one unit cell of SrO. 
The heterostructures were grown at 600 $^{\circ}$C under an oxygen partial pressure of $1\times10^{-5}$ Torr. 
Under this condition, the carriers are considered to be confined in the interfacial region of a few nm thickness \cite{LAOSTO-AFM}. 
XPS measurements were performed using a Scienta SES-100 electron-energy analyzer and a Mg K$\alpha$ ($h\nu = 1253.6$ eV) source. 
Samples were transferred from the PLD chamber to the spectrometer chamber {\it ex situ} and no surface treatment was performed prior to the PES measurements. 
All the measurements were carried out at room temperature with the total energy resolution of about 800 meV. 

\begin{figure}
\begin{center}
\includegraphics[width=\linewidth]{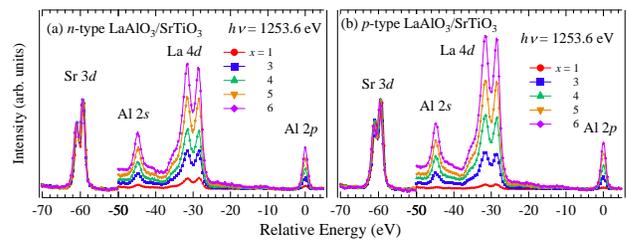}
\caption{(Color online) La, Al, and Sr core-level spectra of the LaAlO$_{3}$($x$ uc)/SrTiO$_3$ (001) samples with varying LaAlO$_{3}$-layer thickness $x$. (a) LaAlO$_3$ layer grown on the TiO$_2$-terminated SrTiO$_3$ substrates forming an $n$-type interface. (b) LaAlO$_3$ layer grown on the SrO-terminated SrTiO$_3$ substrates forming a $p$-type interface. All the spectra have been normalized to the area of the Sr $3d$ core-level spectrum and the Sr $3d$ core-level peak positions have been aligned. }
\label{cores}
\end{center}
\end{figure}
Figure~\ref{cores} shows the La, Al, and Sr core-level spectra of the LaAlO$_3$ thin films grown on the termination-layer-controlled SrTiO$_3$ substrates as described above. 
Because of unavoidable charging effects of the entire sample and resultant uncertainties in the absolute binding energies, the spectra have been aligned to the peak position of the Sr $3d$ core level. 
All the spectra have been normalized to the intensity of the Sr $3d$ core level. 
With increasing LaAlO$_3$ thickness, the intensity of the La and Al core level systematically increases. 

\begin{figure*}
\begin{center}
\includegraphics[width=.6\linewidth]{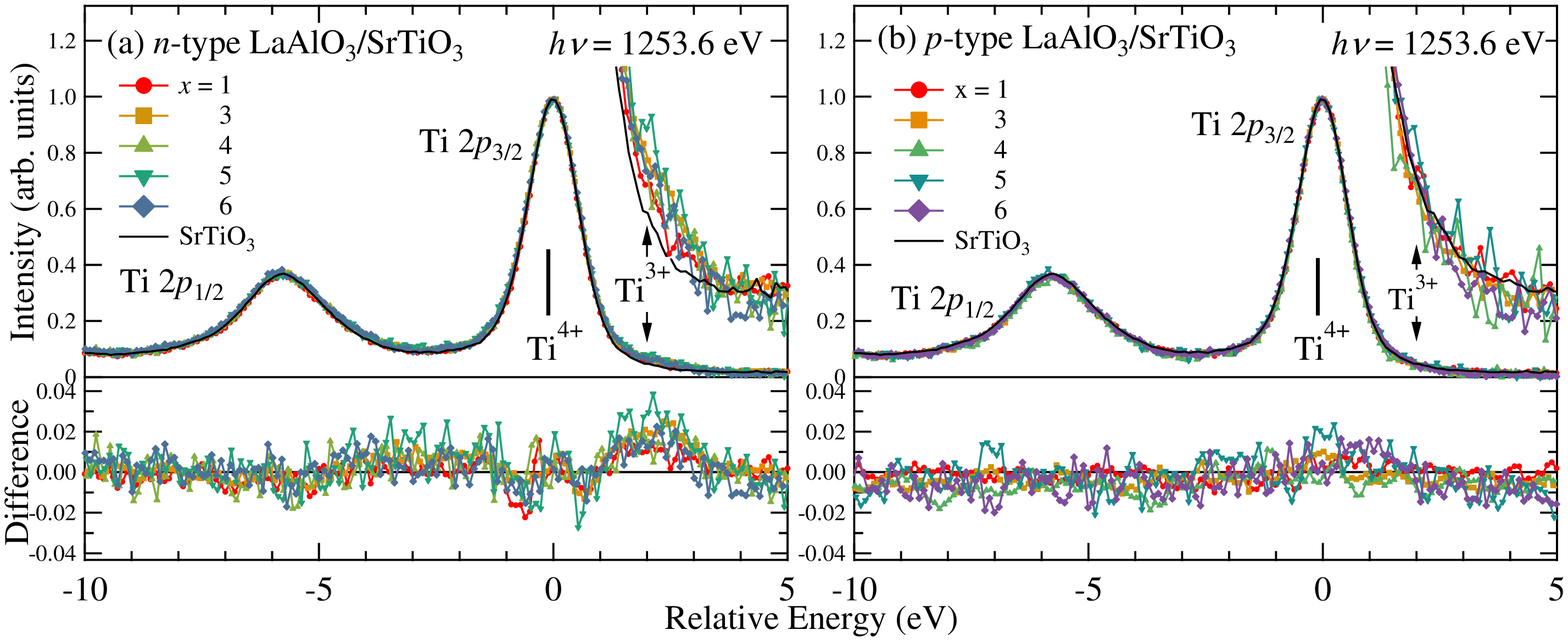}
\includegraphics[width=.3\linewidth]{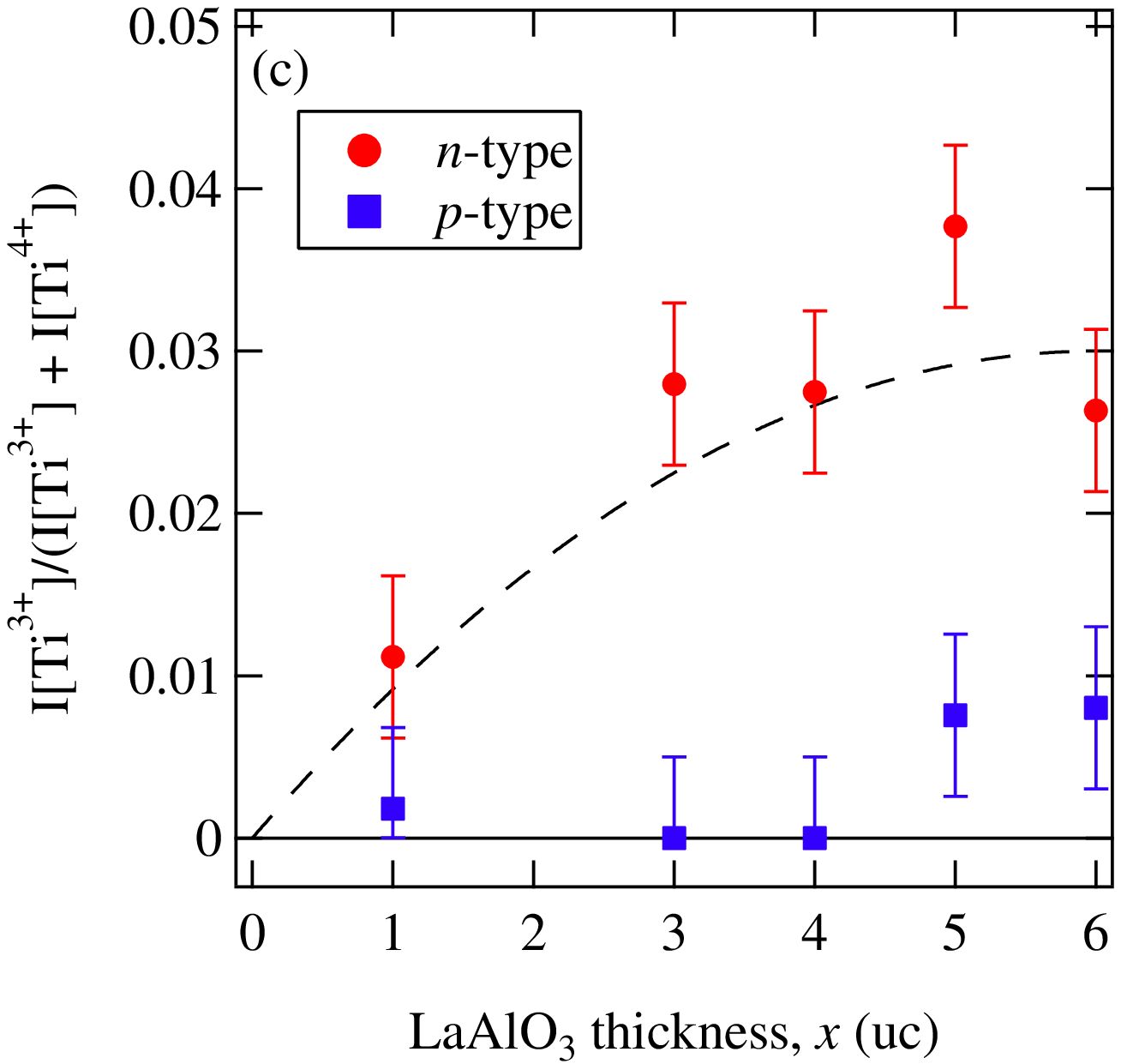}
\caption{(Color online) Ti $2p$ core-level spectra of LaAlO$_{3}$($x$ uc)/SrTiO$_{3}$ (001) samples with varying LaAlO$_{3}$-layer thickness $x$. (a) LaAlO$_{3}$ layers grown on the TiO$_{2}$-terminated SrTiO$_{3}$ substrates forming $n$-type interfaces. The bottom panel shows difference spectra from that of SrTiO$_{3}$. (b) LaAlO$_3$ layers grown on the SrO-terminated SrTiO$_{3}$ substrates forming $p$-type interfaces. (c) Ti$^{3+}$ intensity ratio of the Ti $2p$ core-level spectra as a function of $x$. The dashed curve is a guide to the eye. }
\label{Ti}
\end{center}
\end{figure*}
Figure~\ref{Ti} shows the Ti $2p$ core-level spectra of the same samples. 
The spectra have been aligned to the Ti $2p_{3/2}$ peak position and normalized to its peak height. 
For the LaAlO$_3$ thin films grown on TiO$_2$-terminated SrTiO$_3$ substrates forming $n$-type interfaces [Fig.~\ref{Ti}(a)], one can see an additional feature which can be assigned to Ti$^{3+}$ \cite{XPSYCTO} on the lower binding-energy side of the Ti $2p_{3/2}$ peak and grows with the LaAlO$_3$-layer thickness. 
In order to evaluate the amount of the Ti$^{3+}$ signal, we have subtracted the spectrum of SrTiO$_{3}$ and evaluated the area of the Ti$^{3+}$ components by a standard line-shape fitting. 
Figure~\ref{Ti}(c) shows the resulting Ti$^{3+}$ intensity as a function of the LaAlO$_3$-layer thickness. 
Already for thickness as small as 1 uc, a weak Ti$^{3+}$ component ($\sim 1$ \% of the Ti$^{4+}$ peak) appeared and grew to $\sim 3$ \% for thicknesses greater than 3 uc. 
Combining this Ti$^{3+}$/Ti$^{4+}$ intensity ratio and the mean-free path of photoelectrons of $\sim 2$ nm ($\sim 5$ uc), the total amount of Ti$^{3+}$ per 2D unit cells is estimated to be at least 0.10 - 0.15 \cite{Ti3note}, and the 2D carrier density is estimated to be at least $\sim 10^{14}$ cm$^{-2}$, significantly larger than $\sim 2 \times 10^{13}$ cm$^{-2}$ \cite{ThielLAOSTO} estimated from the transport measurements. 
This indicates that a considerable portion of electrons at the $n$-type interfaces are localized, and are probably responsible for the magnetism of the interfaces \cite{Dikin}. 

Another point to note is that, while the transport data show an abrupt increase of conductivity above the critical LaAlO$_3$-layer thickness of 4 uc, the Ti$^{3+}$ intensity started to increase gradually well below 4 uc. 
This suggests that the electronic reconstruction or charge transfer to the interfacial region occurred already below the critical thickness as reported in a previous work \cite{Ogawa}. 
A similar behavior was also observed in polar multilayers LaAlO$_3$/LaVO$_3$/LaAlO$_3$ \cite{HiguchiLAOLVO, TakizawaLAOLVO}, suggesting that the transport properties of interfaces, which show an abrupt change at 4 uc, are not solely determined by the carrier number but also by the carrier mobility \cite{Bell}, perhaps involving a percolation threshold. 
As the origin of the gradual electronic reconstruction, we consider surface defects and/or adsorbed molecules on the LaAlO$_3$ surface, such as oxygen vacancies which generate and donate electrons to the $n$-type interface. 
A similar mechanism has been discussed by Cen {\it et al}. to explain the creation of metallic regions using a conducting AFM tip \cite{Cen}. 
This mechanism has been taken into account in the theoretical description of the LaAlO$_3$/SrTiO$_3$ interface by Bristowe {\it et al.} \cite{Bristowe}.
According to their calculation, the Ti$^{3+}$ component gradually appears below the critical thickness and reproduces our experimental observation of Ti$^{3+}$ below 4 uc [Fig.~\ref{Ti}(c)]. 
In fact, a recent study has shown that adsorption of polar solvents on the surface of LaAlO$_3$/SrTiO$_3$ samples induces charge carrier at the interface, indicating remote charge transfer from the surface to the interface \cite{Xie}. 
The difference between the amount of Ti$^{3+}$ determined by photoemission and the carrier density determined by the transport measurements below 4 uc indicates that all the carriers in the Ti $3d$ band are localized below 4 uc \cite{Xie}. 

\begin{figure}
\begin{center}
\includegraphics[width=\linewidth]{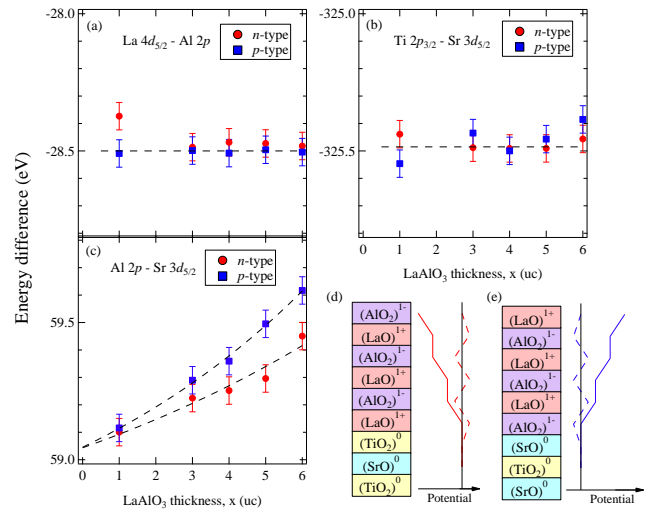}
\caption{(Color online) Shifts of core levels for LaAlO$_{3}$($x$ uc) thin films grown on the SrTiO$_{3}$ substrates. (a) Relative shifts within the LaAlO$_{3}$ layers. (b) Relative shifts within the SrTiO$_{3}$ substrates. (c) Relative shifts between the LaAlO$_{3}$ and SrTiO$_{3}$ layers. (d),(e) Schematic diagram of the Madelung potential in the LaAlO$_3$/SrTiO$_3$ samples with the $n$-type and $p$-type interfaces. The dashed and solid curves represent the electrostatic potential with and without electronic reconstruction, respectively. }
\label{shift}
\end{center}
\end{figure}
We have also studied the core-level shifts as functions of the LaAlO$_3$-layer thickness as shown in Fig.~\ref{shift}. 
Figure~\ref{shift}(a) and (b) indicates that, with increasing LaAlO$_3$ thickness, all the core levels of LaAlO$_3$ are shifted by the same amount [Fig.~\ref{shift}(a)] and those of SrTiO$_3$ are also shifted by the same amount [Fig.~\ref{shift}(b)] (including charging effects), while the core levels of LaAlO$_3$ and those in SrTiO$_3$ are shifted to a different degree [Fig.~\ref{shift}(c)]. 
Considering the fact that the intensity of photoemission signals decreases exponentially with depth, the shifts of the La and Al core levels represent the electrostatic potential at the surface and those of Sr and Ti core levels represent the electrostatic potential at the interface. 
According to the polar catastrophe model, the shift should be as large as $\sim 1$ eV/uc until the critical thickness of 4 uc is reached, where the top of the O $2p$ band at the LaAlO$_{3}$ surface is raised above the bottom of the Ti $3d$ band of SrTiO$_{3}$. 
Figure~\ref{shift}(c) indicates that the experimental core-level shifts of the LaAlO$_3$ layer relative to those of the SrTiO$_3$ were less than $\sim 0.1$ eV/uc. 
Furthermore, the relative shifts were in the same direction for the $n$-type and $p$-type interfaces, in contrast to the expectations of the polar catastrophe model shown in Fig.~\ref{shift}(d) and (e). 
In Ref.~\cite{Segal}, the shifts were also small but the relative shifts between the $n$-type and $p$-type interfaces were opposite to the present result. 
The existence of charged defects at the surface and resulting charge transfer from the surface to the interface explain the reduction of the core-level shifts from $\sim 1$ eV/uc to $\sim 0.1$ eV/uc, but the remaining small shifts may depend on the subtle charge distribution and may vary between the samples prepared by the MBE method \cite{Segal} and the present PLD method. 

In conclusion, we have performed a core-level photoemission spectroscopy study of LaAlO$_{3}$/SrTiO$_{3}$ interfaces as a function of LaAlO$_{3}$-layer thickness. 
Only for LaAlO$_3$ grown on TiO$_2$-terminated SrTiO$_3$ substrates ($n$-type interface), the Ti $2p$ core-level spectra showed a Ti$^{3+}$ component in addition to Ti$^{4+}$. 
The amount of the Ti$^{3+}$ component was significantly larger than the carrier densities estimated from transport measurements, indicating that part of the carriers at the interfaces are localized. 
The gradual rather than abrupt appearance of the Ti$^{3+}$ component with the growth of the LaAlO$_3$ layer as well as the absence of the large core-level shifts expected from the polar catastrophe model suggests that the evolution of charged surface defects or adsorbates needs to be involved beyond the simplest polar catastrophe model \cite{Bristowe}. 

We thank S.~Ishibashi, K.~Terakura, N.~C.~Bristowe, and P.~B.~Littlewood for informative discussions. 
This work was supported by a Grant-in-Aid for Scientific Research from the Japan Society for the Promotion of Science (S22224005) and by the Department of Energy, Office of Basic Energy Sciences, Division of Materials Sciences and Engineering, under Contact No. DE-AC02-76SF00515.


\begin{thebibliography}{10}
\bibitem{Ti3note}
Note that the estimated value of 0.10 - 0.15 Ti$^{3+}$ per 2D uc is a lower bound
  because Ti$^{3+}$ may exist beyond the photoelectron mean-free path.
\end{thebibliography}
\end{document}